\documentclass[twocolumn,english,notitlepage,superscriptaddress,nofootinbib]{revtex4-1}
\usepackage[T1]{fontenc}
\usepackage[latin9]{inputenc}
\usepackage{float}
\usepackage{amsmath}
\usepackage{amssymb}
\usepackage{graphicx}
\usepackage{wasysym}
 \usepackage{color}


\begin{document}
\title{Towards directional force sensing in levitated optomechanics}
\author{A. Pontin}
\affiliation{Department of Physics and Astronomy, University College London, Gower
Street, London WC1E 6BT, United Kingdom}

\author{T.S. Monteiro}
\email{t.monteiro@ucl.ac.uk}
\affiliation{Department of Physics and Astronomy, University College London, Gower
Street, London WC1E 6BT, United Kingdom}

\begin{abstract}
Levitated nanoparticles are being intensively investigated from two different perspectives:  as a potential realisation of macroscopic quantum coherence; and as ultra-sensitive  sensors of force, down to the zeptoNewton level, with a range of various applications, including the search for Dark Matter.  Here we propose that mechanical cross-correlation spectra $S_{xy}(\omega)$ offer a new way to sense the direction of an external force: once detector misalignment errors are minimised, the spectral shape of $S_{xy}(\omega)$ directly indicates the orientation of an external stochastic force, offering something akin to a compass in the $x-y$ plane. We analyse this for detection of microscopic gas currents, but any broad spectrum directed force will suffice, enabling  investigation with laboratory test forces with or without cavities. For a cavity set-up, we analyse misalignment imprecisions between detectors and motional modes due to for example optical back-actions that mask the signature of the directed forces, and show how to suppress them.  Near quantum regimes, we quantify the imprecision due to the $x-y$ correlating effect of quantum shot noise.
\end{abstract}
\maketitle


\section{Introduction}
Experiments on levitated nanoparticles, controlled by cavities as well as active feedback methods~\cite{millen2020optomechanics}, are aimed at  two important goals. The first is towards experimental realisations of quantum coherence, including entanglement and quantum superposition, in systems of mesoscopic or macroscopic size. The second is the prospect of ultra-sensitive detection of forces and accelerations, with applications ranging from fundamental physics, like detection of dark matter, to real world applications.

The two goals are by no means exclusive. Indeed, in the longer term, combining them for quantum limited sensing is itself an overarching aim. Recent advances include cooling to  quantum ground state of the centre of mass motion of a nanoparticle by quantum control~\cite{Magrini2021quantum,Novotny2021quantum}  or via the optical mode of a cavity~\cite{delic2020cooling}. In turn, sensing of forces at the atto or zeptoNewton scale has been investigated~\cite{Geraci2016,hebestreit2018sensing,dark_matter_3} with levitated nanoparticles.

Within this second context, we consider directional force sensing. For instance, if we wish to measure the orientation of a  vector force with unknown components $f_x,f_y$, in principle, we could employ 1D methods: we might independently measure $f_x$ and $f_y$ and compare them. Optomechanical force-displacement sensitivity is well studied in 1D.

\begin{figure*}[ht]
{\includegraphics[width=6.7in]{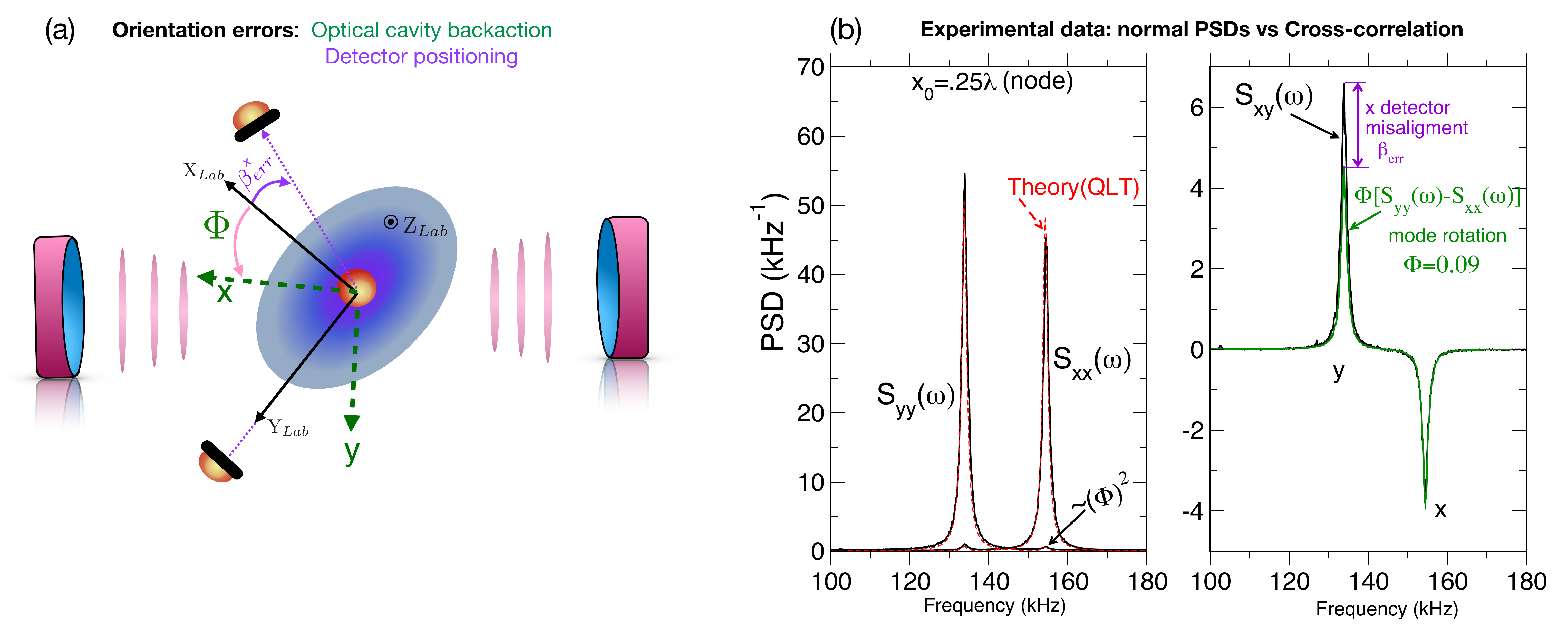}} \caption{ \textbf{(a)} Illustrates two key sources of imprecision in the measurement of orientation of a force in the $x-y$ plane, acting on a nanoparticle levitated  in a trap (blue): (i) the backaction from a surrounding optical cavity, used for cooling, rotates the normal modes (green arrows)  by an angle $\sim \Phi$ away from the laboratory frame $X_L,Y_L\equiv X_{Lab}, Y_{Lab}$ (black arrows), that is set by the tweezer trap. (ii) imprecision $\beta^{x,y}_{err}$ in the alignment between either detector and lab axes. For example, if  either $\Phi\neq 0$ or $\beta^{x}_{err}\neq 0$, measurement of the $x$ motion (in the illustrated example) will acquire an artificial, erroneous $y$ component. \textbf{(b)} Experimental PSDs and cross correlation spectra, $S_{xy}(\omega)$ for a nanoparticle trapped at a cavity node, where there is significant cavity backaction and thus $x-y$ hybridisation.  Data taken from \cite{Pontin2022} that had a $\beta^{x}_{err}= 2^\circ$ offset, while $\beta^{y}_{err}= 0$. The $S_{xy}(\omega)$ PSD clearly shows both these sources of misalignments while for the normal PSDs, however, the signature of hybridisation is hard to detect as it is of order $\Phi^2$. Correcting these misalignments and identifying the $\Phi=0$ trapping point is necessary for our directed force sensing: the departure point is to eliminate unwanted cross-correlations (both $\Phi$ and $ \beta_{err}$), prior to introducing the directed forces.  }
\label{Fig1}
\end{figure*}

\begin{figure*}[ht]
{\includegraphics[width=7in]{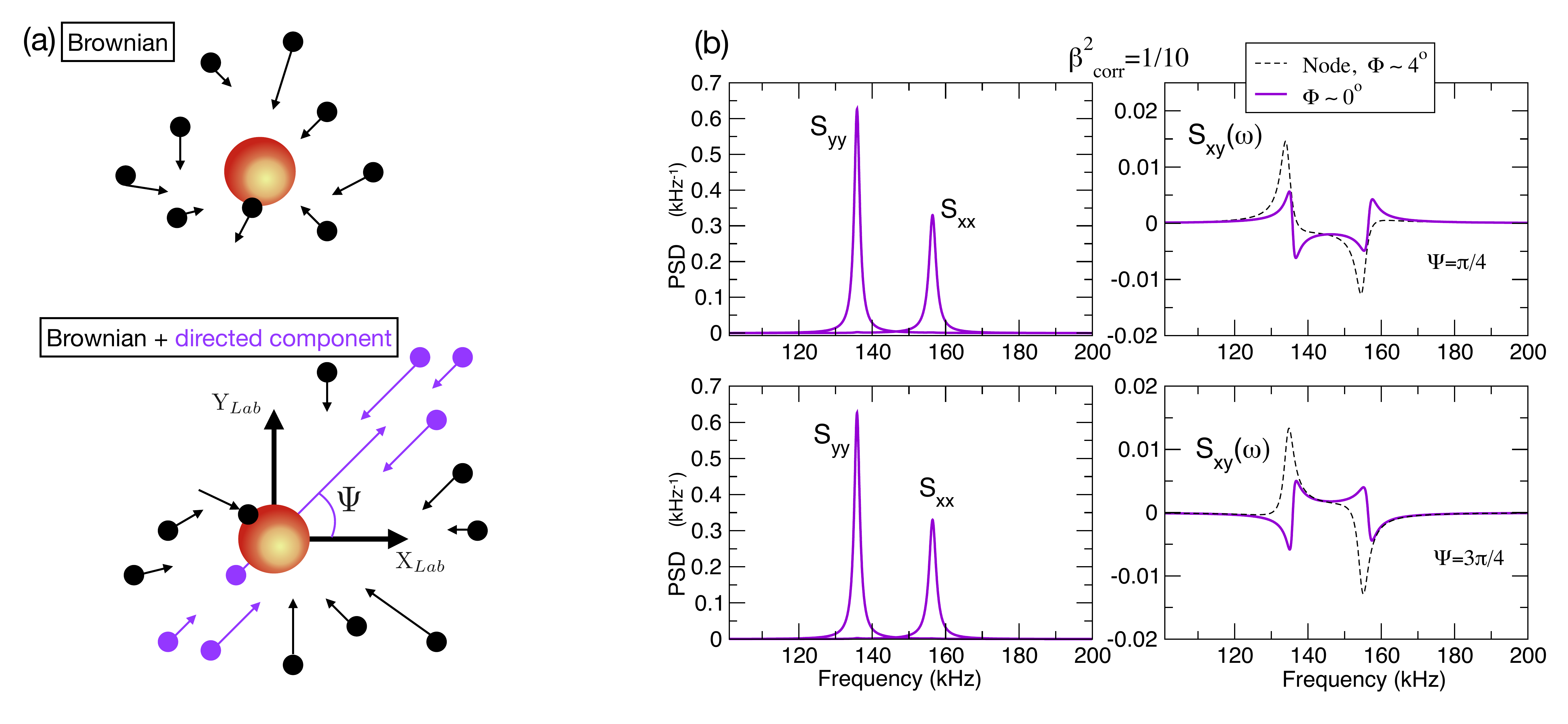}} \caption{ \textbf{(a)} In typical levitated experiments the thermal bath results mainly from collisions with background gas 
which is uncorrelated Brownian motion thus $\langle f^{th}_x(t)  f^{th}_y(t') \rangle=0$ (upper image). Here we allow a directed component, a current of particles at fixed orientation $\Psi$  to $X_{L}$. Although it is still Markovian (in order to model an arbitrary  broad spectrum
external  force)
the $x-y$ motion becomes correlated as  $\langle f^{th}_x(t)  f^{th}_y(t') \rangle \propto \sin \Psi \cos \Psi  \delta(t-t')$.
\textbf{(b)} Shows how we can ``read'' orientation $\Psi$, directly from the shape of $S_{xy}(\omega)$ (right panels). For two different orientations of the directed thermal forces, we plot PSDs and cross-correlation spectra  for the case where the directed component is relatively small ($\beta^2_{corr}=0.1$) .  There is clear dependence on orientation $\Psi$. Pressure $=10^{-4}$\,mbar. Parameters are similar to  experiments with  no directed external force of~\cite{Pontin2022} and are given in the text. 
 If the directed component is small, the cross-correlations induced by the external force can be weaker than those induced by dynamical mode rotations  so are completely masked, for the common procedure of trapping at the node. To expose the external force, the tweezer trap is set to  $x_0=0.145\lambda$ from
 the antinode, the point where the rotation $\Phi$ due to back-action is suppressed (violet lines). This emphasises the importance of eliminating misalignments
 (here by operating with $\Phi \sim 0$), in order to expose the  signature shape of an external directed force. }
\label{Fig2}
\end{figure*}

However, here we propose a different approach which is to measure via the cross-correlation spectrum    $S_{xy} (\omega)= \frac{1}{2} \left(  \langle [\hat{x}]^\dagger \hat{y} \rangle + \langle [ \hat{y}]^\dagger \hat{x}\rangle \right)$. We show it has two very useful features: (1) It offers  greater sensitivity to (and thus ability to correct) misalignments, as errors in the orientation of detectors and normal modes $x,y$ relative to the measurement frame are linear in $S_{xy}$ while they are quadratic in typical PSD spectra $S_{xx},S_{yy}$. This is illustrated in Fig.~\ref{Fig1}. A second important advantage is that (2) the orientation of an external test force is reflected in the spectral line {\em shape} rather than amplitude, allowing a degree of calibration independent measurement. This is illustrated in Fig.~\ref{Fig2} and Fig.~\ref{Fig3}.

The choice of what external force to detect is another important question: studies rely on artificial laboratory test forces, typically electrical, or even gravity $g$, that translate into ultraweak ($\lesssim$  aN) forces acting on the nanoparticle. The question of what microscopic force one might best detect with $S_{xy}$ is also significant. Nanoparticles are also subject to natural thermal forces,  mainly collisions with surrounding gas molecules undergoing Brownian motion. These represent stochastic noise baths of high phonon occupancy $n_B \sim kT/(\hbar\omega_{x,y})$ and are Markovian, so $ \langle f^{th}_{x,y}(t)  f^{th}_{x,y}(t') \rangle \propto \Gamma n_B \delta(t-t')$.  Optomechanical studies  assume that $\langle f^{th}_x(t)  f^{th}_y(t') \rangle=0$, i.e. that the thermal baths acting on each degree of freedom $x,y,z...$ are uncorrelated. 

However, here we introduce a component of stochastic gas collisions with invariant orientation $\Psi$ to the $x$ axis, thus for which $\langle f_x(t)  f_y(t') \rangle \propto \sin \Psi \cos \Psi \delta(t-t')$.  The idea is that a beam of  particles streaming in a particular direction in the $x-y$ plane, although impacting the nanosphere at random, uncorrelated,  times can nevertheless generate $x-y$ correlations. We find its signature in the $S_{xy}$ spectra is a clear and distinctive function of $\Psi$. To our knowledge, this scenario has not previously been considered, but has potential applications both in detection of beams of exotic fundamental particles as well as detection of small gas currents at the high vacuum level.  We emphasize that the directed force need not be  stochastic, but cross-correlation spectra will be more sensitive to broad spectrum $f_x(t)$ and $f_y(t)$, at least on the scale of $|\omega_x-\omega_y|$. We focus on steady state $S_{xy}$, but transients, even single collisions, might also hypothetically  introduce measurable cross correlations, if of sufficient strength.

We focus on levitated cavity optomechanics, although many of the results are generic and apply to cavity-free set-ups. We illustrate our conclusions using the 3D coherent scattering (CS) setup, where a nanoparticle is held in an optical tweezer, but is cooled by a surrounding optical cavity. The CS setup was introduced recently to levitated cavity optomechanics~\cite{delic2019cavity,windey2019cavity} using methods adapted from atomic physics~\cite{vuletic2000laser,vuletic2001three,domokos2002collective,leibrandt2009cavity,hosseini2017cavity}. The novelty is that the cavity is undriven but is populated exclusively by photons coherently scattered by the nanoparticle. The very strong light-matter coupling has already enabled ground-state cooling of the motion along the cavity axis and opened the way to levitated cavity optomechanics at~\cite{delic2020cooling} or near~\cite{Marin2021} quantum regimes.

Although most CS studies have quantum cooled or force-sensed a single degree of freedom, centre of mass motion of a nanoparticle is intrinsically 3D. A recent theoretical analysis of the CS set-up~\cite{MTTM2020,MTTM2021} has shown that hybridisation in the $x-y$ plane is hard to avoid entirely. But the motion naturally decouples into 2+1, with the $z$ motion approximately separate and where one can independently consider hybridisation in the $x-y$ plane. Unlike other optomechanical systems, like membranes, hybridisation for the nanosphere displacements has a clear implication in terms of the directionality of the normal modes. This is important for sensing the direction of a force; indeed, recent work on dark matter detection has also underlined the importance of directional sensing~\cite{dark_matter_1,dark_matter_2,dark_matter_3}.

In section II, we quantify sensitivity to sources of misalignment between our $x,y$ detectors and the laboratory axes $X_L,Y_L$, due to either cavity back-action or experimental imprecision illustrated with data from our recent experimental study~\cite{Pontin2022} of an effective dynamical rotation of normal modes in the $x-y$ plane by an angle $\Phi$. These misalignments mask or degrade the signature of the directional forces we wish to detect, so starting from $\Phi\simeq 0 $ is an essential prerequisite.  In~\cite{Pontin2022} we  showed that there is a cavity trapping position where the rotation angle $\Phi \simeq 0$. We also explain why the condition $\Phi\simeq 0 $ exposes the effect of the directed forces. In section III we introduce our test directional forces: we introduce a white-noise, but directed component  and analyse by solution of quantum Langevin equations. Near quantum regimes we analyse the sensitivity relative to $x-y$ correlations arising from quantum shot noise and characterise the optimal condition as $ C_{xy} \sim 1$ where $C_{xy}= \frac{4g_xg_y}{\kappa\Gamma n_B}$ is a quantum ``cross'' cooperativity. Our key finding is that each direction of the external force is associated with a distinct spectral profile, so independently of the magnitude, the direction can be read-out. The spectral signature is a factor $\sim \Gamma_{opt}/|\omega_x-\omega_y| $ smaller than normal PSDs. For the optical cooling rates $\Gamma_{opt}$ in our set-up~\cite{Pontin2022},
 $\Gamma_{opt}/|\omega_x-\omega_y| \simeq 1/10 $; this can be further optimised, so the effect should be readily detectable experimentally. In Sec IV, we conclude.

\section{Mechanical cross-correlation spectra}

Here we investigate the sensitivity of the cross-correlation to an applied external force. We consider the symmetrised mechanical correlations:

 \begin{equation}
S_{xy} (\omega)= \frac{1}{2} \left(  \langle [\hat{x}]^\dagger \hat{y}\rangle + \langle [ \hat{y}]^\dagger \hat{x}\rangle \right).
\end{equation}

We begin by considering the scenario where the true modes of the motion $ \hat{x},\hat{y}$ are subjected to a rotation, by an angle $\Phi$, in the 2D  $x-y$ plane, relative to the laboratory frame:

\begin{equation}
[\hat{x} \  \hat{y} ]^\top \simeq  R_z(\Phi) [\hat{X}_\text{L} \  \hat{Y}_\text{L}]^\top
\end{equation}

\noindent where $R_z(\Phi)$ is the 2D rotation matrix. For $\Phi$ small, we can write (to quadratic order in the angle):

\begin{equation}
S_{xy} (\omega) \simeq  S^{\text{L}}_{xy} (\omega) +  \Phi[ S_{yy}(\omega) - S_{xx} (\omega)] + \mathcal{O}(\Phi^2) S^{\text{L}}_{xy} (\omega).
\label{CrossMain}
\end{equation}

Above, we defined the cross-correlation spectrum in the laboratory frame as
$S^{\text{L}}_{xy}= \frac{1}{2} \left(  \langle [ \hat{X}_{\text{L}}]^\dagger \hat{Y}_{\text{L}}\rangle + \langle [ \hat{Y}_{\text{L}}]^\dagger \hat{X}_{\text{L}}\rangle \right) $: it is of especial significance here as the effect of the external force will be detected via this term. Hence, if we wish to isolate $S^{\text{L}}_{xy}$ it is important to fully characterise or suppress the terms in $\Phi$.
In particular, the third term would distort $S^{\text{L}}_{xy}$, since if high accuracy is required, the frequency dependence of $\Phi$ cannot be neglected. We show below that rather than a constant angle $\Phi$, a more careful analysis involves a frequency dependent transformation.

{\em In summary:} in the present work, we introduce and consider detection an external force for which $\langle f_x(t)  f_y(t) \rangle\neq 0$. The force induces $x-y$ correlations and produces a measurable effect in $S^{\text{L}}_{xy}$, provided we eliminate or minimise  $\Phi$ and imprecision misalignments, in order to isolate the $S^{\text{L}}_{xy}$ contributions.

\subsection{Orientation  uncertainty due  to misalignment }
In this section we calibrate measurement errors in the absence of the external force, so ${\bf F}(t)=0$ and $S^{\text{L}}_{xy} (\omega)= 0$.

In the direct detection of the motion via scattered light, $x,y$ motions are detected separately as independent time traces peaked around $\omega\simeq \omega_x$ and $\omega_y$ respectively. PSDs in frequency space $S_{xx} (\omega),S_{yy} (\omega)$ and cross-correlation spectra $S_{xy} (\omega)$  are calculated by Fourier transforms of the measured time traces.

There are two types of misalignment between the detectors and the modes of the $x-y$  as illustrated in Fig.~\ref{Fig1}(a) in the main text: (1) dynamical: the optical back action from the cavity and the cavity field induce effective mode rotations $\Phi$ between the true modes and lab frame (2) imprecision from positioning errors $ \beta^{x,y}_{err}$ between the $x,y$ detectors and the lab axes. For small angles, the frame rotation becomes:
\begin{alignat}{1}
\hat{x}(\omega) & \simeq \hat{X}^\text{\text{L}}(\omega)+(\Phi+\beta^x_{err}) \hat{Y}^{\text{\text{L}}}(\omega)\\
\hat{y}(\omega) & \simeq \hat{X}^\text{\text{L}}(\omega)- (\Phi + \beta^y_{err})\hat{Y}^{\text{\text{L}}}(\omega).
\label{1Dto2Derr}
\end{alignat}

Hence

\begin{equation}
S_{xy} (\omega) \simeq   ( \Phi+\beta^x_{err}) S_{yy}(\omega) -( \Phi+\beta^y_{err}) S_{xx} (\omega)
\end{equation}

\noindent Note that a misalignment in the {\em $x$ detector}, appears as a correction of the height of
{\em the $y$  peak} in $S_{xy} (\omega)$ and vice-versa. Specifically, in the experimental traces in Fig.~\ref{Fig1}(b) in the main text, $ \beta^y_{err}\simeq 0$, while $\beta^x_{err}\simeq 0.03$ corresponding to a small misalignment of just $1.7^\circ$ between detector and lab axes. For all the experimental traces in~\cite{Pontin2022}:

\begin{equation}
S_{xy} (\omega) \simeq    (\Phi+ \beta^x_{err}) S_{yy}(\omega) - \Phi S_{xx} (\omega).
\end{equation}

Conversely, in the measured $x$ PSDs, we obtained:
\begin{equation}
S_{xx} (\omega) \to  S_{xx} (\omega\simeq \omega_x) +  (\Phi+ \beta^x_{err})^2 S_{yy}(\omega \simeq \omega_y).
\label{PSDerr}
\end{equation}

The $x$ PSD will gain a small artificial component at $y$ frequency. Whether the error is of dynamical origin or due to a small error in positioning of the detector, the effect will be to incorrectly indicate (or perhaps mask) the effect of an applied external force with components $f_x,f_y$.

As seen in Eq.\ref{PSDerr}, and in the experimental data in Fig.~\ref{Fig1}(b),  in the normal PSDs ($S_{xx},S_{yy}$), the misalignment corrections are of quadratic order and yield very small peaks; thus both types of error are harder to quantify and correct. In contrast, in the $S_{xy}$ cross-correlation spectra, the $\Phi+\beta$ corrections are of linear order and much easier to quantify and correct. As illustrated in Fig.~\ref{Fig1} of the main text, the misalignment of $1.7^\circ$ between detector and lab axes translated to a very substantial change in the height of the $y$ peak, an order of magnitude larger than the $\Phi^2$ feature in the PSD. Thus in future it should be possible to improve alignment of detectors to of order $\lesssim 0.05^\circ$.

 \begin{figure}[ht]
{\includegraphics[width=3.3in]{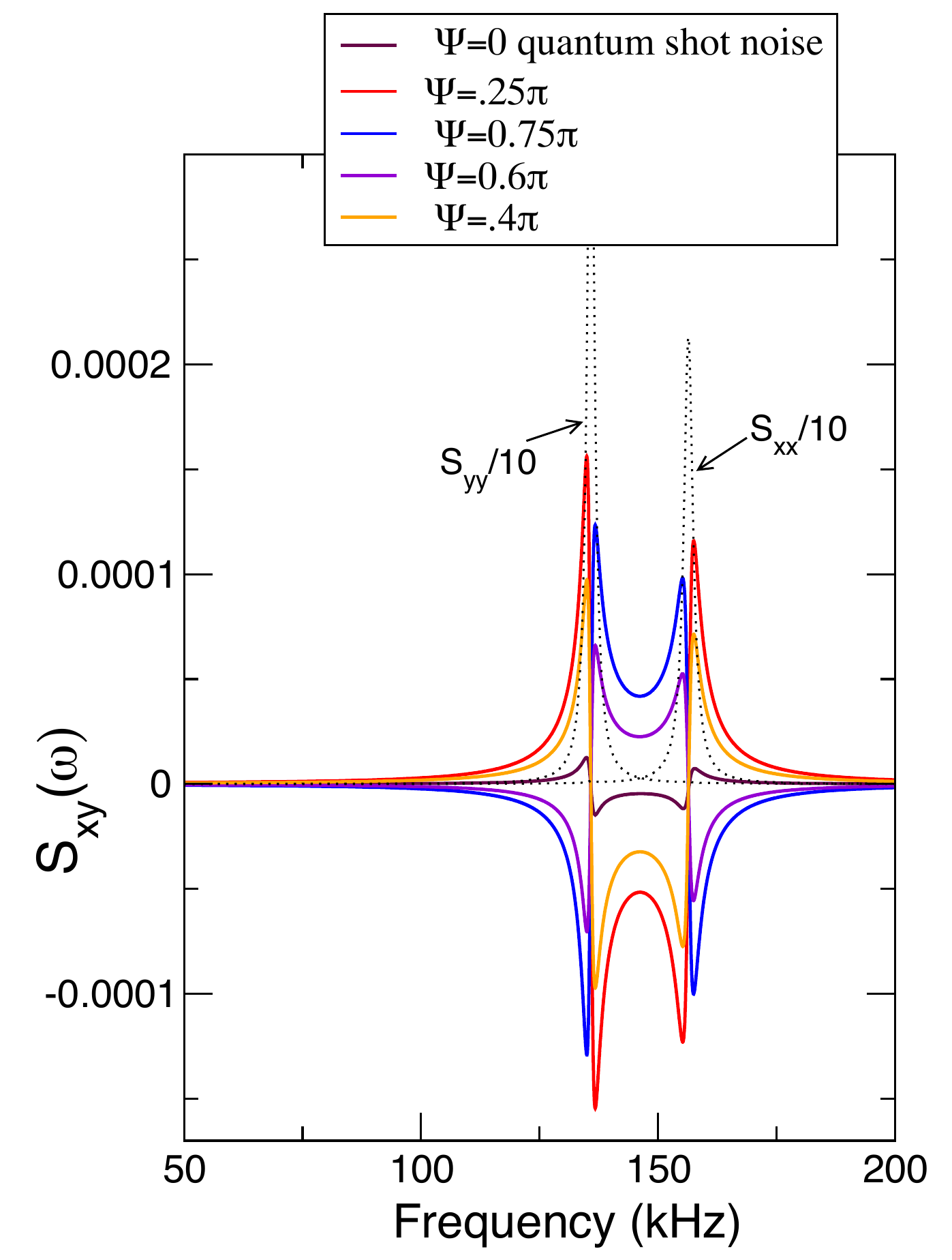}} \caption{Near quantum regimes, with occupancies of a few quanta, we show that the shape of $S_{xy}(\omega)$ depends clearly on $\Psi$ thus the orientation of the external force can be read directly from the line shape. Pressure of the normal Brownian motion $P=5 \times 10^{-7}$\,mbar and for the directed component of comparable strength thus $\beta^2_{corr}=1/4$. For
orientations $\Psi=0, \pi...$ the external directed current does not generate any $x-y$ correlations, but a residual feature is seen: this is due to the optical shot noise (brown line) and is orientation independent. For comparison, the normal PSDs are shown. The correlated feature is  weaker than the $S_{xx},S_{yy}$ by a factor $\sim \Gamma_{opt}/(\omega_x-\omega_y)\sim 1/10$ for the
experiments of~\cite{Pontin2022} but can be optimised. But although the feature is weaker than the PSD, reading a change in shape can be advantageous relative attempting to calibrate to detect a
small change in amplitude of $S_{xx}$ and $S_{yy}$.  }
\label{Fig3}
\end{figure}

The dynamical mode rotation angle $\Phi$ depends sensitively on the trapping position of the particle. In~\cite{Pontin2022} this was varied from node to antinode. $\Phi$ was found to flip sign and to pass through zero in between node and anti-node. For direct detection, Eq.\ref{CrossMain}, the point where $\Phi \simeq 0$ is ideal for detecting correlations induced by an external force.

In a cavity system, one might  also consider detection of cross-correlations via heterodyne detection of the cavity output. In frequency-space, we can write for the cavity mode:
\begin{equation}
\hat{a}(\omega) =\chi_c(\omega) [g_x \hat{x}(\omega) +  g_y \hat{y}(\omega)] +\sqrt{\kappa} \hat{a}_{in}(\omega)
\label{OptMode}
\end{equation}
where the $\chi_c$ is the cavity susceptibility (see below) and $\hat{a}_{in}(\omega)$ are the optical imprecision (ideally quantum shot) noises.  For small $\Phi$, we can approximate:

\begin{eqnarray}
S_{het} (\omega) & \approx   |\chi_c(\omega)|^2 [  (  g_y^2+g_xg_y \Phi) S_{yy}   +  (  g_x^2-g_xg_y \Phi) S_{xx}(\omega) \nonumber \\
     & + g_xg_y  S^{\text{L}}_{xy}(\omega) ] +S_{imp}(\omega)
\end{eqnarray}
where $S_{imp}$ are the imprecision noise contributions  due to incoming shot noise. In this case $\Phi$ appears at linear order and, if positive, increases the $y$ component (that would become the so-called `bright-mode'~\cite{Harris2014,MTTM2021} and decreases the $x$ component (that would become the dark-mode as $\Phi \to \pi/4$. However, the main $x,y$ PSD contributions are always large so heterodyne detection is not as favourable as direct detection in order to isolate and measure $S^{\text{L}}_{xy}$.

\section{Effect of directional forces on measured spectra}
From the previous sections, we wish to detect the effect of an external test force ${\bf F}(t)=f_x(t) {\hat{\bf i}} + f_y(t) {\hat{\bf j}}$ that affects the motions along the lab frame coordinates via its effect on $S^{\text{L}}_{xy}= \frac{1}{2} \left(  \langle [ \hat{X}_{\text{L}}]^\dagger \hat{Y}_{\text{L}}\rangle + \langle [ \hat{Y}_{\text{L}}]^\dagger \hat{X}_{\text{L}}\rangle \right) $.

For a standard optomechanical system, where the degrees of freedom are uncoupled, we can give the solutions of the quantum Langevin equations in terms of thermal and optical quantum noises:
\begin{eqnarray}
\hat{X}^\text{L}(\omega) & = M_{x}^{-1}\left[\sqrt{\Gamma}\tilde{X}^{\text{therm}}+i\sqrt{\kappa}g_{x}\mu_{x}\tilde{A}^{\textrm{in}}\right ] \nonumber \\
\hat{Y}^\text{L}(\omega) & = M_{y}^{-1}\left[\sqrt{\Gamma}\tilde{Y}^{\text{therm}}+i\sqrt{\kappa}g_{y}\mu_{y}\tilde{A}^{\textrm{in}}\right]
\label{1DDisplace1}
\end{eqnarray}
where we have the normalization factor (for $j\equiv x,y$)
$M_{j}(\omega)=1+g_{j}^{2}\mu_{j}(\omega)\eta_c(\omega)$.
The mechanical noise is
\begin{equation}
\tilde{Q}_{j}^{\text{therm}}(\omega)=\chi(\omega,\omega_{j},\Gamma)\hat{b}_{j}^{\text{in}}(\omega)+\chi^{*}(-\omega,\omega_{j},\Gamma)\hat{b}_{j}^{\text{in}}{}^{\dagger}(\omega),
\end{equation}
while the optical noise
\begin{equation}
\tilde{A}^{\textrm{in}}(\omega)=\chi(\omega,-\Delta,\kappa) \hat{a}_{\text{in}}(\omega)+\chi^{*}(-\omega,-\Delta,\kappa)\hat{a}_{\text{in}}^{\dagger}(\omega).
\end{equation}

{\em Optical and mechanical susceptibilities:} The $\mu_{j}(\omega)$ are mechanical susceptibilities, while $\eta_c$ is the optical susceptibility. We have the usual mechanical susceptibility $\mu_j(\omega)=\chi(\omega,\omega_j)-\chi^*(-\omega,\omega_j)$ and optical susceptibility  $\eta_c(\omega)=\chi(\omega,-\Delta)-\chi^*(-\omega,-\Delta)$, where, e.g., $\chi(\omega,\omega_x)=[-i(\omega-\omega_x)+\frac{\Gamma}{2}]^{-1}$ and $\chi(\omega,\Delta)=[-i(\omega-\Delta)+\frac{\kappa}{2}]^{-1}$.

For the normal levitated optomechanics scenario, the thermal noises are dominated by collisions with  surrounding gas molecules. The associated Brownian motion is isotropic, so one assumes:
\begin{equation}
  \langle  [\hat{b}_x^{\text{therm}}]^\dagger \hat{b}_y^{\text{therm}}\rangle =\langle  [\hat{b}_y^{\text{therm}}]^\dagger \hat{b}_x^{\text{therm}}\rangle  =0
\label{Uncorr}
\end{equation}
while $\langle [\hat{b}_x^{\text{therm}}]^\dagger \hat{b}_x^{\text{therm}}\rangle= \bar{n}_x\delta(t-t')$, where $\bar{n}_x=kT/(\hbar \omega_x)$ where $\bar{n}_x$ is the thermal occupancy of the $x$ mode. Similarly for $y$.

Here we lift that assumption by introducing an additional component of gas moving along a definite direction, hence we replace $ \Gamma\hat{b}_{j}^{\text{in}}(\omega)\to  \Gamma  \hat{b}_{j}^{\text{in}}(\omega)+ \Gamma_{j,corr} \hat{b}_{j}^{\text{corr}}(\omega)$. The additional component of correlated noises no longer obeys Eq.\ref{Uncorr} but rather, $ \langle  [\hat{b}_x^{\text{corr}}]^\dagger(t) \hat{b}_y^{\text{corr}}(t')\rangle  \propto \delta(t-t')$.

We model the correlated collisions by a physically intuitive model: a force $F(t)$, at $\Psi=45^\circ$  to $x$ implies that $ f_x(t)=f_y(t)$ at arbitrary times; conversely, at $\Psi=135^\circ$  to $x$
implies that $ f_x(t)=-f_y(t)$ at arbitrary times etc. We consider a broad spectrum force in the $\omega \approx \omega_x,\omega_y$ range, and for simplicity take white noise, so expect $ \langle  f_x (t) f_y(t')\rangle  \propto \sin \Psi \cos \Psi \delta(t-t')$.

For simplicity we consider our directed component to be of the same species as the main Brownian gas collisions but represents a modest fraction $\beta^2_{corr} <1 $ of the gas (it is no problem to relax this assumption). Hence we take $\sqrt{ \Gamma_{x,corr}} \sqrt{\Gamma_{y,corr}} =\Gamma \beta^2_{corr} \sin \Psi \cos \Psi$. 

For the solution of the quantum Langevin equations (labelled quantum linear theory or QLT here) with this  model we obtain:

\begin{equation}
  S^{\text{L}}_{xy} (\omega) =S_{QN}(\omega)+ S_{f_x f_y}(\omega) \ \textrm{where}
\label{Corr2}
\end{equation}
\begin{eqnarray}
  S_{QN}(\omega)&=& \kappa g_xg_y |\chi_c(\omega)|^2 \mathcal{M}_{xy}(\omega) \ \textrm{and}  \nonumber \\
  S_{f_xf_y}(\omega)&=&   \Gamma \frac{\beta^2_{corr}}{2} \sin \Psi \cos \Psi [ (\bar{n}_x+1+ \bar{n}_y+1)\mathcal{M}_{xy}(\omega) \nonumber \\
   & + & (\bar{n}_x+ \bar{n}_y)\mathcal{M}^*_{xy}(-\omega) ]
\end{eqnarray}
and where
\begin{equation}
 \mathcal{M}_{xy}(\omega)=\left[ \frac{\mu_x(\omega)\mu_y^*(\omega)}{M_x(\omega)M_y^*(\omega)} + cc\right]
\end{equation}

In Fig.~\ref{Fig2} we  investigate the effect of introducing a small directed component,$\beta^2_{corr} =1/10 $. Pressure $P=1 \times 10^{-4}$\,mbar. Otherwise, Fig.~\ref{Fig2} numerics employ cavity and tweezer parameters used in the experiments reported in~\cite{Pontin2022}. Pressure $P=1 \times 10^{-3}$\,mbar, tweezer polarisation angle $\theta=49^\circ$, $\Delta= -2\pi \times 176$\,KHz, $\kappa/2= 2\pi\times200$\,kHz, input power $P_{in}=0.485$\,W, nanoparticle radius $R=60.1$\,nm. We note Fig.~\ref{Fig1} used an experimental data point at $\Delta= -2\pi \times 360$\,KHz.

The full cross-correlation $S_{xy}(\omega)$ is calculated and plotted, but since  the tweezer trap is set  at $x_0=0.145\lambda$ away from the  antinode, and this corresponds to the cancellation point, where $\Phi \simeq 0$, as verified experimentally in~\cite{Pontin2022}, the distinctive $S^{\text{L}}_{xy}$ signature line shape is isolated as shown by the violet lines. In contrast, if the tweezer trap is at the usual trapping point, the node of the cavity standing wave, the characteristic correlated line shape is completely masked by the effect of non-zero $\Phi$. 

We expect the above model to be reasonably representative of the effects of  broad range of forces (including different species of gas and forces that are not necessarily Markovian, as long as they are sufficiently broad to span both $x,y$ frequencies) on the cross-correlation spectra.

\subsection{ Sensitivity of force detection relative to quantum shot noise imprecision }
The first term in Eq.\ref{Corr2} relates  to dynamics in a cavity as it represents the effect of incoming quantum shot noise. The quantum shot noise term couples simultaneously to both $x$ and $y$ modes (with relative strengths $g_x$ and $g_y$) and induces cross-correlations between them. The second and third terms are new and represent the effect of the directed forces. But all terms are modulated equally by the frequency dependent function $\mathcal{M}_{xy}(\omega)$. Thus one can estimate the relative strengths; for $-\Delta \sim \omega_x,\omega_y$, $|\chi_c(\omega)|^2\sim 4/\kappa^2$. Hence

\begin{equation}
 S_{QN}(\omega): S_{f_x f_y}(\omega) \sim \frac{4g_xg_y}{\kappa}:\Gamma \beta^2_{corr} \sin \Psi \cos \Psi n_B
\end{equation}
where $ n_B= (\bar{n}_x+ \bar{n}_y)/2$. Thus defining a quantum ``cross-cooperativity'' $C_{xy}$ in close analogy to the usual quantum cooperativity:
\begin{equation}
 C_{xy}= \frac{4g_xg_y}{\kappa\Gamma \bar{n}}
\end{equation}
we see that the external force will dominate the correlating effects of quantum shot noise if
\begin{equation}
 C_{xy} \lesssim \beta^2_{corr} \sin \Psi \cos \Psi.
\end{equation}

In Fig.~\ref{Fig3} we plot $S_{xy}(\omega)$  for the same parameters as Fig.~\ref{Fig2}, except now pressure $P=10^{-6}$\,mbar and $\beta^2_{corr}=1/4$, so the nanoparticle has phonon occupancies are of order $n \approx 5-10$ quanta. For strong quantum cooling a sideband resolved set-up would be required (by using higher finesse mirrors). The behavior is plotted as a function of $\Psi$. For $\Psi=0,\pi$ the only correlations arise from the optical quantum shot noise (the additive $S_{QN}$ contribution), which for these parameters is still small relative to the directed force terms. It is possible to have $\beta^2_{corr} \sin \Psi \cos \Psi \sim 1$ ; and as the force term depends on $\Psi$ its effect may still be clearly detectable even if of the same order as $S_{QN}$.

\subsection{Sensitivity of force detection via cross-correlations relative to usual PSDs }
It is useful to quantitatively estimate the sensitivity of  detection employing $S_{xy}(\omega)$ and $S^{\text{L}}_{xy}$ relative to the ``gold-standard''  of detection of a resonant force by a normal PSD. Intuitively for appreciable cross-correlations, there should be some overlap between the $x,y$ spectral   features, while retaining $x-y$ resolution, thus the width of the spectra should be of the order of $|\omega_x-\omega_y|$ thus $\Gamma_{opt} \sim |\omega_x-\omega_y|$. This necessitates reasonable optical damping since $\Gamma \ll \Gamma_{opt} \sim \Gamma_{tot}$ is small at high vacuum pressures of levitated optomechanics.

For the usual PSD, $S_{xx}$ for example, instead of $\text{Re}[{\mathcal{M}_{xy}(\omega)}]$ function the thermal force terms are weighted by a frequency envelope  $| \mu_x(\omega)|^2/|M_x(\omega)|^2$ which has a maximum for $\omega \approx \omega_x$, i.e.,  $| \mu_x(\omega_x)|^2/|M_x(\omega_x)|^2\sim 1/\Gamma^2_{opt}$ (and similar maximum at $\omega \simeq \omega_y$).

In contrast, for $S^{\text{L}}_{xy} (\omega)$, the envelope has a maximum  $\text{Re}[{\mathcal{M}_{xy}(\omega= \omega_x)}] \sim \frac{1}{\Gamma_{opt}(\omega_x-\omega_y)}$.
Hence, we estimate
\begin{equation}
\frac{S_{xy}(\omega=\omega_x)}{ S_{xx}(\omega=\omega_x)} \sim  \frac{\beta^2_{corr}\sin \Psi\cos \Psi  \Gamma_{opt}}  {\omega_x-\omega_y}
\end{equation}\\
Indeed for the cavity set-up in \cite{Pontin2022},  $\frac{\Gamma_{opt}}{|\omega_x-\omega_y|} \sim 1/10$ and this is consistent with the behavior shown in Fig.~\ref{Fig3}.

\section{Conclusions}
We have shown that mechanical cross-correlations offer new possibilities for sensing of external forces at both the classical and quantum scale: they can indicate the direction of a force directly via the shape of the spectral feature. It opens the way to detect new of forces of microscopic origin such as tiny gas currents, small anisotropies in the background Brownian motion, possibly due to incoming streams of gas; or such as anisotropies due multi-temperature baths~\cite{Millen2014}, but  on far smaller scales here.  

It  offers a new approach in the search for Dark Matter: were there such a `DM wind'  component,  comprising fluxes of directed particles that individually deliver undetectably weak impacts, but at sufficient rate to affect the steady state correlations,  $S_{xy}$ spectral shapes would provide an effective means to  characterise it. This scenario is not compatible with current paradigms that suggest DM particles are detectable via individual, rare recoils with nuclei; steady state $S_{xy}$ correlations require interactions on the timescale of the damping timescale (eg $\Gamma^{-1}_{tot} \sim 0.1-1$ ms for quantum cavity cooled nanospheres). However,  the $S_{xy}$ spectra  might, with very little added effort, exclude the presence of such correlations of  fundamental  origin,  as part of current experiments. In addition, even for temporal traces of  single-recoil events, mechanical cross correlations complement current techniques.

Our focus here has been on sensing with cavities: the main role of the cavity is to provide optical damping so that the condition $\Gamma_{opt} \sim |\omega_x-\omega_y|$ is met. In addition, following a study that investigated how to achieve $\Phi\simeq 0$ in a cavity set-up, the suppression of misalignment errors that mask $S^{\text{L}}_{xy}$ is well understood. But otherwise, the results are generic: work is underway to investigate $S_{xy}$ with a directional test force in an active (feedback) cooling set-up.  It is expected that photon recoil from $z-$aligned tweezer photons, like normal gas Brownian motion, cannot generate $x-y$ correlations.  An important goal will be to characterise and minimise sources of classical or quantum noise that can generate  $x-y$ correlations (as distinct from heating) that might affect the
characteristic  $S^{\text{L}}_{xy}$ signal.

{\em Acknowledgements} The authors would like to acknowledge helpful discussions with Markus~Rademacher,  Fiona~Alder  and Marko~Toro\v{s}. We acknowledge funding from the Engineering and Physical Sciences Research Council (EPSRC) Grant No. EP/N031105/1.
\bibliographystyle{unsrt}
\bibliography{2Dbiblio}

\newpage
\begin{widetext}
\section*{Appendix}
\subsection{ Hybridisation of $x-y$ motion in coherent scattering set-ups}

We briefly review the CS set-up. The dynamics has been introduced elsewhere ~\cite{delic2019cavity,windey2019cavity,delic2020cooling} and we have investigated
3D hybridisation \cite{MTTM2020,MTTM2021,Pontin2022}. While the cavity dynamics is not central to the directed force sensing it is  an effective mechanism to quantum cool and in particular to
achieve the sensing condition $\Gamma_{cool} \simeq |\omega_x-\omega_y|$. In addition, in \cite{Pontin2022}, a  means to achieve $\Phi\simeq 0$ and good alignment, essential to expose the $x-y$ correlations generated by the external force,  was demonstrated. In principle, a different set-up,
even without a cavity,  might identify the orientation of the normal modes
and seek to accurately align detectors at those orientations. But the $\Phi \simeq 0$ cancellation points in \cite{Pontin2022} are robust to even fluctuations in experimental parameters
so are advantageous.

\subsection{Equivalence to frame rotation relative to  lab frame ${X}_\text{L}, {Y}_\text{L}$}

Typical cavity cooling set-ups currently trap a nanoparticle in an optical tweezer that, here, we take to have a principal axis
 along the $z$ axis. An additional optical cavity is then used  for quantum cooling of  the 3D
centre of mass coordinates $x,y,z$. Using the coherent-scattering set-up where the cavity is populated only
by tweezer photons coherently scattered by the nanoparticle, one can
  decouple the dynamics into a 2+1 system: the $z$ frequency is far from resonance so one can accurately consider only $x-y$ motion.

Hence although our calculations are always fully 3 dimensional, we analyse and discuss only motion in the 2D $x-y$ plane.
For future sensing of a directional vector  force ${\bf F}(t)$, we must define a fixed laboratory frame. It is convenient to define this as the fixed axes of the tweezer trap . Thus we equate
$[\hat{x}^\text{1D} \ \hat{y}^\text{1D}]^\top \equiv [\hat{X}_\text{L} \  \hat{Y}_\text{L}]^\top$.\\
Then, in the reduced 2D space, a solution of the Langevin equations describes the $x-y$ hybridisation in the form
\cite{MTTM2020,MTTM2021}:

\begin{alignat}{1}
\hat{x}(\omega) & = \hat{x}^\text{1D}(\omega)+\mathcal{R}_{xy}(\omega)\hat{y}(\omega)\\
\hat{y}(\omega) & = \hat{y}^\text{1D}(\omega)+\mathcal{R}_{yx}(\omega)\hat{x}(\omega).
\end{alignat}

Above, $\hat{x}(\omega),\hat{y}(\omega)$ denote the spectra of the true modes of the motion;
$ \hat{x}^\text{1D}(\omega), \hat{y}^\text{1D}(\omega)$ denote the motion along the axes of the tweezer trap that also define our laboratory frame. The
$\mathcal{R}_{xy}(\omega),\mathcal{R}_{yx}(\omega)$ denote hybridisation functions (discussed below)
that characterise $x-y$ mixing, due to dynamical effects such as backaction of the cavity optical mode as well as additional co-trapping by the cavity standing wave.

One can further re-arrange and write the modes in terms of the unperturbed modes of the tweezer:
\begin{alignat}{1}
\hat{x}(\omega) & = \mathcal{N}^{-1}(\omega) [\hat{x}^\text{1D}(\omega)+\mathcal{R}_{xy}(\omega)\hat{y}^{\text{1D}}(\omega)]\\
\hat{y}(\omega) & = \mathcal{N}^{-1}(\omega) [\hat{y}^\text{1D}(\omega)+\mathcal{R}_{yx}(\omega)\hat{x}^{\text{1D}}(\omega)].
\label{1Dto2D}
\end{alignat}
where $\mathcal{N}(\omega)= 1- \mathcal{R}_{xy}(\omega)\mathcal{R}_{yx}(\omega)$.

From Eq.\ref{1Dto2D}  we write:

\begin{equation}
S_{xy} (\omega) \simeq  S^{\text{L}}_{xy} (\omega) +  \text{Re}({\mathcal{R}_{yx}}) S_{xx}(\omega)  +\text{Re}({\mathcal{R}_{xy}}) S_{yy} (\omega) + {\mathcal{R}_{xy}}{\mathcal{R}^*_{yx}}\langle [ \hat{y}^{1D}]^\dagger \hat{x}^{1D}\rangle + {\mathcal{R}_{xy}}{\mathcal{R}^*_{yx}}\langle [ \hat{x}^{1D}]^\dagger \hat{y}^{1D}\rangle
\end{equation}
where we retained terms up to quadratic order in $\mathcal{R}_{xy}, {\mathcal{R}_{yx}}$ (noting that the full exact 3D solutions are of cubic order and it is is a reasonable assumption to consider scenarios where the  hybridisation function is not large.

It was shown in \cite{Pontin2022})   that the  hybridisation functions $ \text{Re}{\mathcal{R}_{xy}} \approx -\text{Re}{\mathcal{R}_{yx}} \simeq \Phi$,  where it was experimentally shown that the $x-y$ modes
are in general  anti-correlated, a consequence of the opposing signs  of $ \text{Re}{\mathcal{R}_{xy}}$ and $\text{Re}{\mathcal{R}_{yx}}$. Further details and explicit forms for the
$\mathcal{R}_{xy}, {\mathcal{R}_{yx}}$ for the cavity set-up are given below. But using $ \text{Re}{\mathcal{R}_{xy}} \approx -\text{Re}{\mathcal{R}_{yx}} \simeq \Phi$ recovers Eq.\ref{CrossMain}
$S_{xy} (\omega) \simeq  S^{\text{L}}_{xy} (\omega) +  \Phi[ S_{yy}(\omega) - S_{xx} (\omega)] + \mathcal{O}(\Phi^2) S^{\text{L}}_{xy} $.

The approximation of the hybridisation functions by a constant $\Phi$ is exact at
 in the limit of large detunings ($ -\Delta \gg \omega_x, \omega_y$ and small $\Phi$). However, it remains a useful model, even near resonant regimes $ -\Delta \sim \omega_x \sim \omega_y$. We always perform exact calculations on PSD spectra, without neglecting the frequency dependence of $\Phi$. or making assumptions on the size of $\Phi$, but the rotation model has proved remarkably accurate and insightful in analysis of experiments \cite{Pontin2022}.

{\em In summary}, the effect of the interaction between the nanoparticle and the cavity dynamics and cavity field
results in $x-y$ hybridisation. This can be modelled by a rotation  of the nanoparticle modes in the $x-y$ plane, by an angle $\Phi$ relative to the fixed laboratory frame.

\subsection{$\Phi$ for coherent scattering tweezer-cavity system}

{\em Physical model}: In a coherent scattering (CS) approach, the optical cavity is not externally driven but it is populated exclusively by light scattered by the nanoparticle. The corresponding Hamiltonian results from the coherent interference between the electric fields of the tweezer and cavity  $\hat{H}=-\frac{\alpha}{2}\vert\mathbf{\hat{E}}_{\text{cav}}+\mathbf{\hat{E}}{}_{\text{tw}}\vert^{2},$ where  $\alpha$ is the polarizability of the nanosphere.  The interference term $\propto (\mathbf{\hat{E}}_{\text{cav}}^{\dagger}\mathbf{\mathbf{\hat{E}}}_{\text{tw}}+\mathbf{\hat{E}}_{\text{cav}}\mathbf{\mathbf{\hat{E}}}_{\text{tw}}^{\dagger})$ gives rise to an effective
 CS potential:
 \begin{equation}
\hat{V}_{\text{CS}}/\hbar = -E_{d}\text{cos}(\phi+k \hat{Y}_c) e^{-({\hat{x}^{2}/w_{x}^{2} +\hat{y}^{2}/w_{y}^{2} })} \left[\hat{a}+\hat{a}^{\dagger}\right].
\label{eq:VCS}
\end{equation}

 In  ~\cite{delic2019cavity,windey2019cavity,delic2020cooling}, a linearised
effective Hamiltonian  was derived and presented, with conservative terms:

\begin{equation}
\frac{\hat{H}_1}{\hbar}= -\Delta \hat{a}^\dagger\hat{a} + \sum_k \omega_k \hat{b}_k^\dagger\hat{b}_k
 + \sum_k g_k (\hat{b}_k^\dagger+ \hat{b}_k)(\hat{a}^\dagger+ \hat{a})
\label{Hnode}
\end{equation}
 for $k\equiv x,y$ and neglecting terms in $z$.
Later it was augmented \cite{MTTM2020} to include direct coupling terms, still within the linearised framework:
\begin{equation}
\frac{\hat{H}}{\hbar}=\frac{\hat{H}_1}{\hbar}+ g_{xy} \hat{x} \hat{y}
\label{Htot}
\end{equation}

And from the above, for the  specific combined CS  tweezer trap plus cavity set-up, the hybridisation functions can be given explicitly \cite{MTTM2020,MTTM2021}:
\begin{equation}
\mathcal{R}_{xy}(\omega) = \frac{i\mu_{x}(\omega)}{M_{x}(\omega)} G(\omega)  \ \textrm{and} \  \mathcal{R}_{yx}(\omega)= \frac{i\mu_{y}(\omega)}{M_{y}(\omega)} G(\omega)
\end{equation}

where $G(\omega)= \left[i\eta_{c}(\omega)g_{x}g_{y}+g_{xy}\right]$ is a term that represents the interference between the
`direct' static coupling  between $x$ and $y$ (proportional to $ g_{xy}$); and an indirect, cavity mediated, coupling term (
proportional to $g_{x} g_{y}$). The prefactors $M_{j}(\omega)=1+g_{j}^{2}\mu_{j}(\omega)\eta_c(\omega)$ include a small
optical backaction correction to each displacement.
For our simplified analysis, we take $M_{j} \simeq 1$.  Numerical tests showed this is
an excellent approximation. The reason for this is that the small backaction correction
 is peaked around each of the mechanical frequencies, i.e. at $M_x(\omega \approx \omega_x)$, and $M_y(\omega \approx \omega_y)$
 while for the cross-correlation, we show below the values around $M_x(\omega \approx \omega_y)\approx 1$, and $M_y(\omega \approx \omega_x) \approx 1$
are most important.

 We note that, in the present discussion, we  refer to both the cavity mediated couplings $\eta_{c}(\omega)g_{x}g_{y}$
as well as the usual optomechanical back-action terms $g_j^2 \eta_0(\omega)$ as  `optical backaction' terms, but clearly, in the former case, the optical backaction acts on different mechanical modes.

Using the above expressions, we can readily show (see \cite{Pontin2022}) that in the absence
of an external force  i.e for $S^{1D}_{xy}=0$, then:

\begin{equation}
S_{xy} (\omega) \approx  \frac{G(\omega)}{\omega_x-\omega_y} [  S_{yy}(\omega) - S_{xx}(\omega) ]\label{Sxyapprox}
\end{equation}

Since $G(\omega \sim \omega_x) \simeq G(\omega \sim \omega_y) \equiv G$, is approximately constant, taking $\Phi \sim  \frac{G}{\omega_x-\omega_y}$, we recover Eq.\ref{PSDerr} (without the detector
misalignment):
we can write:
\begin{equation}
S_{xy} (\omega) \approx \Phi [S_{yy}(\omega) - S_{xx}(\omega) ]\label{Sxyexpt}.
\end{equation}

The above expression was used in \cite{Pontin2022} to measure $\Phi$ over a range of experimental parameters, by showing that $S_{xy} (\omega)$ is accurately equivalent to a rescaled $S_{yy}(\omega) - S_{xx}(\omega)$ (in the absence of an external directed force).

\subsection{Suppression of $\Phi$}

  If the term $G(\omega) \left[i\eta_{c}(\omega)g_{x}g_{y}+g_{xy}\right] \simeq 0$, the destructive interference between $x-y$ coupling and indirect, cavity-mode mediated coupling suppresses hybridisation and hence
$S_{xy} \simeq 0$.

Since the direct coupling $g_{xy}\simeq-g_{x}g_{y}\frac{2\text{Re}(\bar{\alpha})\cos{\phi}}{E_{d}\sin^{2}{\phi}}$, and $\bar{\alpha}\simeq -iE_{d}\text{cos}(\phi)[i\Delta-\kappa/2]^{-1}$:
\begin{equation}
g_{xy}\simeq g_{x}g_{y}\left[\frac{2\Delta\cot^{2}{\phi}}{\Delta^{2}+\frac{\kappa^{2}}{4}}\right].\label{gjk}
\end{equation}
Thus depending on the positioning, $\Delta$ or $\kappa$, the direct
couplings contribution can be similar or exceed the cavity mediated
coupling. Direct and indirect contributions, in general,interfere
destructively. We can show that $i\eta_c(\omega)\to\frac{-2\Delta}{(\kappa/2)^{2}+\Delta^{2}}$
if $\text{-}\Delta\gg\omega$ (and we are interested primarily in
the region $\omega\sim\omega_{j}$). Thus for large $-\Delta$:

\begin{equation}
G(\omega)\approx g_{x}g_{y}\left[\frac{-2\Delta}{\Delta^{2}+(\kappa/2)^{2}}\right]\left[1-\cot^{2}\phi\right],\label{xycouple}
\end{equation}
and we see that the $G$ is real and frequency independent. Furthermore, at $\phi=\pi/4$  the $x$-$y$ hybridisation almost fully vanishes so we have a cancellation point where the $S_{xy}$ correlation
spectra are near zero. Thus for large $-\Delta$, the cancellation point requires trapping at a  point $x_0=0.125 \lambda$, thus midway between node and antinode.

The above was derived in \cite{MTTM2020} and experimentally demonstrated in \cite{Pontin2022}. In the latter work it was also found that
at the cancellation point the mode  frequencies return to their unperturbed values (thus optical spring corrections are eliminated). It was also found that
as $-\Delta \to \omega_{x,y}$, the cancellation point moves towards the node. Thus for the calculations of Figs.~\ref{Fig2} and~\ref{Fig3}, for $\Delta=-176$ kHz
for the cancellation point, we used   $x_0=0.145 \lambda$.

There can be strong three-way hybridisation  of the normal modes in the $x-y$ plane  with optical modes,  in strong-coupling regimes. For $g_{x}\approx g_{y}$, which is obtained experimentally for tweezer polarisation angles near  $\theta=\pi/4$,  the strong-coupling regime  can lead to the formation of dark and bright modes~\cite{Harris2014,MTTM2021}. While hybridisation phenomena are seen in other set-ups (e.g. membranes),  for the levitated  optomechanics here, they represent
 a change in orientation of the motional modes, so are  significant for applications such as directional force sensing.

\end{widetext}
\end{document}